\documentstyle[12pt,dina4p]{article}
\newcommand{\be}{\begin{equation}}
\newcommand{\ee}{\end{equation}}
\newcommand{\ba}{\begin{eqnarray}}
\newcommand{\ea}{\end{eqnarray}}
\newcommand{\bc}{\begin{center}}
\newcommand{\ec}{\end{center}}

\newcommand{\bay}{\begin{array}{rcl}}
\newcommand{\eay}{\end{array}}
\newcommand{\dis}{\displaystyle}
\newcommand{\text}{\textstyle}

\newcommand{\rf}[1]{(\ref{#1})}

\renewcommand{\arraystretch}{2.0}
  \renewcommand{\theequation}{\thesection.\arabic{equation}}
  \newcommand{\mysection}[1]{\section{#1}\setcounter{figure}{0}
                    \setcounter{table}{0}\setcounter{equation}{0}}
\def\Tr{\mbox{\rm Tr}}
\def\tr{\mbox{\rm tr}}

\def\DD{D\kern-0.82em\lower-.2ex\hbox{$\not$}\kern+0.8em}

\def\th{\theta}

\def\Gk{\Gamma_k}
\def\gb{\bar{g}}

\def\fs{
   \mbox{$ F_{\mu\nu}^a \kern -1.65em  \lower -.2ex \hbox{${}^*$}$}
     \kern +1.5em
       }
\def\fsn{
   \mbox{$ F \kern -0.9em  \lower -.2ex \hbox{${}^*$}$}
     \kern +0.35em
       }
\def\ff{F_{\mu\nu}^a F_{\mu\nu}^a}
\def\ffs{F_{\mu\nu}^a \fs}
\def\ffsn{F \fsn}

\begin{document}
\noindent
\begin{titlepage}
      \begin{flushright}
        DESY 96--065\\
        hep-th/9604124
      \end{flushright}
  \vspace{1.0in} \LARGE
  \begin{center}
    {\bf Renormalization of the Topological Charge in Yang--Mills Theory
     }\\ \vspace{0.3in}
  \end{center}
  \vspace{0.2in} \large
  \begin{center}
    {\bf {M. Reuter} \\
     \it {Deutsches Elektronen-Synchrotron DESY\\
     Notkestrasse 85  \\
     D-22603 Hamburg   \\
     Germany}
     }\\ \vspace{0.3in}
  \end{center}
  \vspace{1.0in}
 \normalsize

\begin{abstract}
The conditions leading to a nontrivial renormalization of the
topological charge in four--dimensional Yang--Mills theory are
discussed. It is shown that if the topological term is regarded
as the limit of a certain nontopological interaction, quantum
effects due to the gauge bosons lead to a finite multiplicative
renormalization of the $\th$--parameter while fermions give
rise to an
additional shift of $\th$. A truncated form of an
exact renormalization group equation is used to study
the scale dependence of the $\th$--parameter. Possible
implications for the strong CP--problem of QCD are discussed.
\end{abstract}

\end{titlepage}

\mysection{Introduction}
One of the most interesting aspects of Yang--Mills theories
in 4 spacetime dimensions is the possibility of adding
a term $S_{\rm top}= i\th Q$ to their action which
is proportional to the topological charge $Q$ :
\be\label{1.0}
S_{\rm top}[A]\equiv i\th\frac{\bar{g}^2}{32\pi^2}
           \int d^4x F_{\mu\nu}^a \fs
\ee
From a hamiltonian point of view the vacuum angle $\th$
 can be regarded as a kind of quasi--momentum which
owes its existence to the periodic structure of the
Yang--Mills vacuum and which is similar to the
quasi--momentum of Bloch waves in periodic potentials.
For this reason it was commonly believed that $\th$  is not renormalized
by radiative corrections, and that all
observables are $2\pi$--periodic in $\th$.
It came as a surprise therefore that explicit one--loop
calculations \cite{shif, anjoh}
within standard lagrangian perturbation theory
revealed a finite renormalization of the topological charge.
Later on it was observed \cite{joh} that the zero--modes of the inverse
gluon propagator also lead to a renormalization of the
topological charge and that their contribution cancels
precisely the finite renormalization found earlier \cite{shif, anjoh}.
Even though there seems to be no net renormalization
left the cancellation which leads to this result is of
a rather delicate nature. The first one of the two contributions
has the character of a triangle anomaly and
originates in the ultraviolet while the second one, due
to the zero--modes, is a typical infrared effect.
As the cancellation has been established at the one--loop
level only one might wonder if it persists at
higher orders of perturbation theory and at the non--perturbative
level. Since the infrared behavior of
QCD--type theories is only very poorly understood one
cannot exclude the possibility that the actual contribution of
the zero--modes differs from the lowest order result
and that the compensation is incomplete therefore.

In this paper we shall explain in which sense
one may talk about a renormalization of the
topological charge or of the $\th$--parameter, and how
this can be reconciled with the hamiltonian
non--renormalization argument. Both the infrared
and the ultraviolet effects will be investigated in
detail, and we shall see that generically there is
no perfect compensation among them.
Because we are aiming at a clean separation of
the relevant momentum scales, we employ the method
of the exact renormalization group equations \cite{rge}.
The basic idea is to consider a scale--dependent
effective action $\Gamma_k$, henceforth referred to as the ``effective
average action", which obtains from the classical action $S$
by integrating out only the field modes with momenta
larger than the infrared cutoff $k$. The conventional
effective action $\Gamma$ is recovered in the limit $k\to0$, i.e.,
in the space of all actions, the renormalization
group trajectory $\Gamma_k$, $0\leq k< \infty$, interpolates between
the classical action $S=\Gamma_{k\to\infty}$ and the standard effective
action $\Gamma=\Gamma_{k\to0}$.

In ref.\cite{rw3} we introduced an exact evolution
equation for gauge theories which maintains gauge
invariance at all intermediate scales.\footnote
{For alternative approaches see \cite{warr,bon,ell}.}
For a pure Yang--Mills theory it reads
\be\label{1.1}
\kern-0.5em
\begin{array}{rcl}
\dis
k\frac{d}{dk}
\Gamma_k
[A,\bar{A}]
&=&\dis
\frac{1}{2}\Tr
  \left[
    \left(
 \Gamma_k^{(2)}[A,\bar{A}]
         +R_k\left(\Delta[\bar{A}]\right)
    \right)^{-1}
   k\frac{d}{dk}
         R_k\left(\Delta[\bar{A}]\right)
  \right] \\
&-&\dis \Tr
      \left[
         \left(
              -D_\mu[A] D_\mu[\bar{A}]
              +R_k\left(-D^2[\bar{A}]\right)
         \right)^{-1}
         k\frac{d}{dk}R_k\left(-D^2[\bar{A}]\right)
      \right]

\eay
\ee
As we use the background gauge fixing technique \cite{abb},
the functional $\Gamma_k$ depends both on the usual classical
average field $A_\mu^a$ and on the background field
$\bar{A}_\mu^a$.
The equation (\ref{1.1}) has to be solved for the initial
condition
\be\label{1.2}
\Gamma_\infty [A,\bar{A}]
=
S[A] +\frac{1}{2\alpha}\int d^4x
\left(
   D_\mu^{ab}[\bar{A}](A_\mu^b-\bar{A}_\mu^b)
\right)^2
\ee
Apart from the classical action $S[A]$, $\Gamma_\infty$
also contains
the well--known background gauge fixing term \cite{abb}.
$\Gamma_k^{(2)}[A,\bar{A}]$ denotes the matrix of second
functional
derivatives of $\Gamma_k$ with respect to $A$, at fixed
$\bar{A}$.
The function $R_k$ describes the precise form of the
infrared cutoff. It is arbitrary to a large extent, but
it has to satisfy $\dis\lim_{u\to\infty} R_k(u)=0$ and
$\dis\lim_{u\to0} R_k(u)=Z_k k^2$ for some constant $Z_k$
(see below).
Usually we shall use the parametrization
\be\label{1.3}
R_k(\Delta)=Z_k k^2 R^{(0)}
\left(\frac{\Delta}{Z_k k^2}\right)
\ee
with $R^{(0)}$ smoothly interpolating between $R^{(0)}(0)=1$
and $R^{(0)}(\infty)=0$.
The operator $\Delta$ is used to distinguish, in a gauge
invariant way, ``high momentum" modes from ``low momentum"
modes. Expanding all field modes in terms of eigenfunctions
of $\Delta$, only the modes with eigenvalues $p^2>k^2$ are
integrated out. In practice $\Delta$ consists essentially of (minus)
the covariant Laplacian $-D^2[\bar{A}]$ with the covariant
derivatives in the adjoint representation. The meaning of
the factor $Z_k$ in (\ref{1.3}) is as follows. Assume that, at
scale $k$, a certain field mode has a massless inverse
propagator $Z^\prime_k p^2$. Then we should use
$Z_k\equiv Z_k^\prime$
in (\ref{1.3}) because this guarantees that the inverse
propagator and the cutoff combine to $Z_k(p^2+k^2)$
for small eigenvalues $p^2$ of $\Delta$. Hence the mode is
cut off at $p^2 \approx k^2$ by a kind of field--dependent mass term.
Actually $Z_k$ may be chosen differently for different
types of fields. It also may depend on $\bar{A}$, but
not on $A$. (See \cite{rw3,rw5} for further details and \cite{corfu}
for a review of this approach.)

Approximate but still nonperturbative solutions of
eq. (\ref{1.1}) with (\ref{1.2}) can be obtained by truncating
the space of all action functionals. If one makes an
ansatz for $\Gamma_k$ with finitely many $k$--dependent
parameters
(generalized couplings) multiplying the field monomials
which were retained, then the functional evolution equation
becomes a set of ordinary differential equations for the
generalized couplings. To be precise, in writing down eq.
(\ref{1.1}) we already made a special kind of truncation. As
it stands the evolution equation neglects renormalization
effects in the gauge fixing and the ghost
sector.\footnote{A detailed discussion of this approximation
and the general form of the evolution equation can be found
in refs. \cite{rw5,corfu}.}
In the cases studied so far this has led to rather reliable
results \cite{rw3,ah,num}.

The objective of this paper is to solve the
renormalization group equation with the initial
condition
\be\label{1.4}
S[A]=\frac{1}{4}\int d^4x
\ff
+ i\th_{\rm bare}
     \frac{\bar{g}^2}{32\pi^2}\int d^4x
\ffs
\ee
We are going to allow for a scale-dependent $\th$--parameter,
$\th\equiv\th(k)$, and we shall follow its evolution
from the bare value $\th_{\rm bare}\equiv\th(\infty)$ down to the
renormalized one, $\th_{\rm ren}\equiv\th(0)$.\footnote{
For a different evolution equation in the framework of the
dilute instanton gas approximation see also ref.\cite{km}.}
The classical action (\ref{1.4}) coincides
with the effective action $\Gamma_\Lambda$ at the UV cutoff
$\Lambda\to\infty$.
Let us see what happens if we lower the scale $k$
from $\Lambda$ to an infinitesimally lower scale
$\Lambda-\delta k$.
Near $k=\Lambda$ the Hessian $\Gamma_k^{(2)}$ which appears on the
RHS of (\ref{1.1}) is simply $\Gamma_\Lambda^{(2)}=S^{(2)}$. However, as
a consequence of the topological nature of the $\th$--term,
its matrix of second functional derivatives vanishes
identically, and $S^{(2)}$ receives contributions only
from the standard kinetic term $\frac{1}{4}F^2$. Therefore
$\Gamma_\Lambda^{(2)}$
contains no parity-odd piece. This entails that
the traces in (\ref{1.1}) cannot produce a term proportional
to the pseudoscalar $\ffs$ which could match a term
\be
k \frac{d\th(k)}{dk}\int d^4x \ffs
\ee
on the LHS of the equation. Hence
$\frac{d\th}{dk}=0$ at $k=\Lambda$, and
$\th(\Lambda-\delta k)=\th(\Lambda)$ remains
unchanged. Though the parity--even terms in
$\Gamma_k$ have changed while going from $\Lambda$ to
$\Lambda-\delta k$,
we can repeat the above argument for the full
range of scales between $k=\Lambda$ and $k=0$. The result
is that $\th(k)$ keeps its bare value $\th(\Lambda)$ at
all lower scales, i.e., it does not get
renormalized.\footnote{ For the general form of the evolution equation
\cite{rw5} this is still true.}

Within the renormalization group formalism, the above argument is the
analog of the hamiltonian reasoning
which leads to the conclusion that $\th$ is not
renormalized. The crucial question is how this can
be reconciled with the explicit diagrammatic calculations
in ref.\cite{shif} which yield a finite renormalization
of the topological charge or, equivalently, of the $\th$--parameter.
In our framework this phenomenon
can be explained as follows. Let us temporarily replace
the topological term in (\ref{1.4})  by
\be\label{1.5}
S_\th[A,\phi] = i\th_{\rm bare} \frac{\bar{g}^2}{32\pi^2}
                \int d^4x \,\phi(x) \ffs
\ee
Here $\phi(x)$ is a localized external pseudoscalar field which
we shall not quantize. We interpret the term (\ref{1.5})
as the coupling of a pseudoscalar ``meson" $\phi(x)$
to the gluon field with a bare coupling strength
$\th_{\rm bare}=\th(\Lambda)$. If we now ask how the coupling
$\th=\th(k)$ depends on the scale $k$ we indeed will get a
nontrivial answer. The second variation of
(\ref{1.5}) is no longer zero, but rather proportional to
$\int d^4x\partial_\mu\phi K_\mu^{(2)}$ where $K_\mu$ is the
Chern--Simons current. Therefore the $k$--evolution produces all
sorts of terms involving both $\phi$ and $A_\mu^a$. Among them there
is the term $\th(k) \int d^4x\phi \ffs$ with a scale dependent
coupling $\th(k)$. After having solved the evolution
equation for the renormalization group trajectory
$\Gamma_k[A,\bar{A}; \phi]$ we can ask what happens if
we allow $\phi(x)$ to approach unity for all $x$.
Then, on the one hand, (\ref{1.5}) is the original
topological term again, but on the other hand also the running
interaction term $\th(k)\int\ffs$ becomes proportional to the
topological charge {\it but with a renormalized prefactor $\th(k)$}.
Later on we shall demonstrate that -- if understood in this sense --
a renormalization of the topological charge is indeed possible.

The situation is most concisely described by saying that
the $k$--evolution and the limit $\phi(x)\to 1$ do not
commute. If one sets $\phi(x)\equiv 1$ from the outset
the topological charge is not renormalized in accordance
with the hamiltonian arguments. If one considers the topological
term as the limit of the interaction term
$\int \phi \ffs$ for the slowly varying $\phi$ but lets
$\phi(x)\to1$ only at the end of all calculations then one finds
a nontrivial renormalization of $\th$. Clearly the
two different procedures correspond to different
physical situations; which is the correct one
cannot be decided on purely formal grounds. In the
following we shall study the second option throughout.

The remaining sections of the paper are organized as follows.
In \mbox{section 2} we derive and solve the evolution
equation of $\th(k)$ for all non--zero values of $k$.
We establish that $\th(k)$ has a finite discontinuity
at $k\to\infty$ and is constant otherwise. In \mbox{section 3}
we investigate the limit $k\to 0$ and demonstrate
that $\th(k)$ has a second discontinuity at $k=0$.
In section 4 we summarize our results and
comment on possible applications in the context of the
strong CP problem of QCD. In the main part of this paper
we discuss the more interesting effects due to the quantized
gauge field.
In the appendix we include fermion loops, and the reader
should compare the respective calculations for gauge bosons
and fermions.

\newpage

\mysection{Ultraviolet Renormalization}

We consider pure Yang--Mills theory with an arbitrary (semisimple, compact)
gauge group $G$ in 4--dimensional euclidean space.
In order to solve the evolution equation we make an ansatz of the
following form\footnote{We write $\bar{g}$ for the bare gauge coupling and
$D_\mu^{ab}[A]=\partial_\mu\delta^{ab}-i\bar{g}A_\mu^c(T^c)^{ab}$
with \mbox{$(T^c)^{ab}=-if^{cab}$} for the covariant derivative in the adjoint
representation. Furthermore, \mbox{
$\fs\equiv\frac{1}{2}\varepsilon_{\mu\nu\alpha\beta}F_{\alpha\beta}^a$ }
with $\varepsilon_{1234}=1$.}

\be\label{2.1}
\begin{array}{rcl}
\dis \Gk [A,\bar{A};\phi]& =& \dis Z_F(k)\int d^4x
  \left\{
 \frac{1}{4}F_{\mu\nu}^a(A)F_{\mu\nu}^a(A)
        +\frac{1}{2\alpha}
           \left(D_\mu^{ab}[\bar{A}](A_\mu^b-\bar{A}_\mu^b)\right)^2
  \right\} \\
 &+&\dis i\th(k)\frac{\gb^2}{32\pi^2}
    \int d^4x \,\phi(x)\,F_{\mu\nu}^a(A)\,\fs(A)
\end{array}
\ee
It satisfies the initial condition (\ref{1.2}) with the classical
action (\ref{1.4}) for the values $Z_F(\infty)=1$ and
$\th(\infty)=\th_{\rm bare}$. The ansatz (\ref{2.1}) truncates the space
of all actions to a 2--dimensional subspace
parametrized by $Z_F$ and $\th$. If one inserts (\ref{2.1}) into the
evolution equation (\ref{1.1}) one obtains ordinary differential
equations for the funtions $Z_F(k)$ and $\th(k)$.
In order to fully specify the evolution equation one has to make a
choice for the cutoff operator $\Delta[\bar{A}]$.
As in refs.\cite{rw3,rw5} we take
\be\label{2.1a}
\Delta[\bar{A}]=\Gk^{(2)}[A=\bar{A},\bar{A};\phi=0]
\ee
but at the level of physical quantities neither the precise
definition of $\Delta$ nor that of $R^{(0)}$ will matter.

In the approximation used here
corrections to the gauge fixing
term  are neglected, and therefore the background field
 $\bar{A}$ enters (\ref{2.1})
only via the classical gauge fixing term. This means
that it is sufficient for our purposes to know
$\Gk[A,\bar{A}=A;\phi]$ because from the $F^2$-- and the
$F\fsn$--term we can read off $Z_F(k)$ and $\th(k)$,
respectively. For $\bar{A}=A$ the LHS of (\ref{1.1}) reads
\be\label{2.2}
\bay
\dis k\frac{d}{dk} \Gk [A,A;\phi] & = & \dis
\frac{1}{4} k \frac{dZ_F(k)}{dk}\int d^4x \ff \\
&+&\dis i k \frac{d\th(k)}{dk} \frac{\gb^2}{32\pi^2}
   \int d^4x \,\phi(x)\ffs
\eay
\ee
From now on all field strengths and covariant derivatives are
constructed from $A$. In evaluating the RHS
of the evolution equation we have to recall that
$\Gk^{(2)}\equiv\Gk^{(2)}[A,\bar{A};\phi]$ is the matrix of second
derivatives with respect to $A$ only; hence these derivatives
have to be performed {\it before} one sets $\bar{A}=A$.
Keeping this in mind one arrives at
\be\label{2.3}
\Gk^{(2)}[A,A;\phi] =Z_F(k) {\cal D} +i\th(k)\frac{\gb^2}{8\pi^2}
                    {\cal V}
\ee
with the operators
\be\label{2.4}
\bay
\dis {\cal D}_{\mu\nu}^{ab} &=&\dis
                           -D_\alpha^{ac}D_\alpha^{cb}\delta_{\mu\nu}
                           +2i\bar{g} F_{\mu\nu}^{ab}
                           +(1-\alpha^{-1})D_\mu^{ac}D_\nu^{cb} \\
\dis {\cal V}_{\mu\nu}^{ab} & = &
      \dis \varepsilon_{\mu\nu\alpha\beta}(\partial_\alpha \phi)
           D_\beta^{ab}
\eay
\ee
Here $F_{\mu\nu}^{ab}\equiv F_{\mu\nu}^c (T^c)^{ab}$ is the field
strength matrix in the adjoint representation. When one inserts
(\ref{2.3}) into \rf{1.1} consistency requires us to retain only the
first terms of the derivative expansion. Because
$\Gk[A,A;\phi]$ is a gauge invariant functional of $A$,
the lowest order terms are proportional to $\ff$ and $\phi \ffs$.
These are the same field monomials as on the LHS, eq. (\ref{2.2}),
so we can compare their coefficients and deduce the differential
equations for $Z_F(k)$ and $\th(k)$. Expanding
\be\label{2.5}
\bay
\dis (\Gk^{(2)}[A,A;\phi]+R_k)^{-1}&=&
 \dis (Z_F(k){\cal D}+ R_k)^{-1} \\

 \dis -i\th(k)\frac{\gb^2}{8\pi^2}
     (Z_F(k){\cal D}+R_k)^{-1}&\dis {\cal V}&
     \dis (Z_F(k) {\cal D}+R_k)^{-1}
 +{\cal O}(\phi^2)
\eay
\ee
the term independent of $\phi$ contains no $\varepsilon$--tensor
and gives rise to the $\ff$--invariant, whereas the term
linear in $\phi$ contributes to $\ffs$. Thus we get the following
decoupled equations
\be\label{2.6}
\kern-.5em
\begin{array}{rl}
\dis\frac{1}{4}k \frac{d}{dk} Z_F(k) \int d^4x \ff &=
\dis
\frac{1}{2} \Tr
   \left[
     \Big(Z_F(k){\cal D}+R_k(\Delta)\Big)^{-1}
     k\frac{d}{dk}R_k(\Delta)
   \right]
\\
&
\dis
\!\!
-\Tr\dis
      \left[
        \left(-D^2+R_k(-D^2)\right)^{-1}
        k\frac{d}{dk}R_k(-D^2)
      \right] + \cdots
\eay
\ee
\be\label{2.7}
\kern-.5em
\begin{array}{l}
\dis
k\frac{d}{dk}\th(k)\int d^4x \phi(x)\ffs =
\\
\dis
 -2\th(k)\Tr
   \left[
      \Big(Z_F(k){\cal D}+R_k(\Delta)\Big)^{-1}
      {\cal V}
      \Big(Z_F(k){\cal D}+R_k(\Delta)\Big)^{-1}
k\frac{d}{dk}R_k(\Delta)\right] + \cdots
\eay
\ee
Our goal is to extract the pieces proportional to
$F^2$ and $\phi \ffsn$ from the traces \rf{2.6} and \rf{2.7}.
To do this we may insert any field configuration
into the traces which discriminates unambiguously
between the respective invariants. Because of the
complicated operators involved, this procedure is by far
more convenient than the standard derivative expansion
techniques. We shall specify $A_\mu^a$ later on. For the time
being let us only assume that for the gauge field chosen, the
Yang--Mills equations $D_{\mu}^{ab}F_{\mu\nu}^b\equiv 0$
are satisfied. This has the following very useful consequence.
If one defines the operators ${\cal D}_T$ and $D\otimes D$ by
\be\label{2.8}
\bay
\dis
({\cal D}_T)_{\mu\nu}^{ab} & = &
\dis
\left( -D^2 \delta_{\mu\nu}+ 2i\bar{g} F_{\mu\nu}\right)^{ab}\\
\dis
\left(D\otimes D\right)_{\mu\nu}^{ab} & = &
\dis
D_\mu^{ac} D_\nu^{cb}
\eay
\ee
then $[{\cal D}_T,D\otimes D]=0$ for such fields.
This implies that the operators
\be\label{2.9}
P_L=-\frac{D\otimes D}{{\cal D}_T} \qquad, \qquad
P_T=1-P_L
\ee
are orthogonal projectors on generalized longitudinal
and transverse gluon states in the background $A$:
$P_{T,L}^2 =P_{T,L}$, $P_TP_L=0$. The kinetic operator ${\cal D}$
decomposes according to
\be\label{2.10}
{\cal D}= \left[P_T+\frac{1}{\alpha}P_L\right] {\cal D}_T
\ee
Leaving the ${\cal V}$--term aside, the inverse propagator
for the transverse and the longitudinal modes is $Z_F(k){\cal D}_T$
and $\alpha^{-1}Z_F(k){\cal D}_T$, respectively. Moreover,
after setting $\bar{A}=A$, the cutoff operator \rf{2.1a} becomes
$\Delta=Z_F(k){\cal D}$. In view of the comments following
eq.\rf{1.3} this suggests the following choice for the
factors $Z_k$ entering the cutoff function
$R_k$\footnote{A priori the $Z_k$'s are defined in terms of
$P_T[\bar{A}]$ and $P_L[\bar{A}]$.}:
\be\label{2.11}
Z_k=\left[P_T+\frac{1}{\alpha}P_L\right]Z_F(k)
\ee
Hence \rf{1.3} becomes
\be\label{2.12}
R_k(\Delta)=\left\{1-\left(1-\alpha^{-1}\right)P_L\right\}
            Z_F(k) k^2R^{(0)}\left({\cal D}_T/k^2\right)
\ee
This form of $R_k$ has to be used in the first trace
on the RHS of \rf{2.6} and in the one of \rf{2.7}, since
these traces are due to the gauge boson fluctuations.
The second trace in \rf{2.6} stems from the
Faddeev--Popov ghosts. Because the renormalization of their
kinetic term is neglected, one simply sets $Z_k=1$
there \cite{rw3,rw5}.

The equation \rf{2.6} for $Z_F(k)$ has been evaluated
in ref.\cite{rw3} already and we only quote the result
here. The renormalized gauge coupling constant is defined by
\be\label{2.13}
g^2(k)=\bar{g}^2Z_F(k)^{-1}
\ee
Its $\beta$--function is
\be\label{2.14}
\beta_{g^2}=k\frac{d}{dk}g^2(k)= g^2(k)\eta_F(k)
\ee
where
\be\label{2.15}
\eta_F(k)\equiv - k\frac{d}{dk} \ln Z_F(k)
\ee
denotes the anomalous dimension of the gauge field.
Eq.\rf{2.6} leads to the following $\beta$--function
\be\label{2.16}
\beta_{g^2}=-\frac{11 T(G)}{24\pi^2}g^4
            \left[
              1-\frac{5T(G)}{24\pi^2}g^2
            \right]^{-1}
\ee
where $T(G)$ denotes the value of the quadratic Casimir
operator in the adjoint representation:
$f^{acd}f^{bcd}=T(G)\delta^{ab}$.
With our conventions one has $T(G)=N$ for $G=SU(N)$.
Eq.\rf{2.16} is a
nonperturbative result. It sums up contributions of all
orders in $g^2$. Expanding for small $g^2$, the $g^4$--term
coincides with the standard one--loop expression, and
the $g^6$--term differs by only a few percent from the
known 2--loop coefficient.

Turning now to the equation for $\th(k)$, \rf{2.7}, the
properties of $P_L$ and $P_T$ can be used to simplify it
considerably:
\be\label{2.17}
k\frac{d}{dk}\th(k)
\int d^4x \phi(x)\ffs
=
2\th(k)Z_F(k)^{-1}\left\{T_1+T_2\right\}
\ee
with
\be\label{2.18}
\bay
\dis T_1 & \dis \equiv &\dis
\Tr\left[{\cal V} \left\{1+(\alpha-1)P_L\right\}
k\frac{d}{dk}
\left({\cal D}_T+k^2R^{(0)}({\cal D}_T/k^2)\right)^{-1}\right]
\\
\dis T_2 & \dis \equiv &\dis
\dis
\eta_F(k)k^2\,\Tr \Bigg[ {\cal V}\left\{1+(\alpha-1)P_L\right\}
                        R^{(0)}({\cal D}_T/k^2)
\\
&&\dis\qquad\qquad\qquad
\cdot
\left({\cal D}_T+k^2R^{(0)}({\cal D}_T/k^2)\right)^{-2}
\Bigg]
\eay
\ee
The contribution $T_1$ is similar to what one encounters
in a  one--loop calculation with an IR cutoff, whereas the
second term, $T_2$, contains the ``renormalization group
improvement". The factor $\eta_F(k)$ arises when the
$k$--derivative acts upon the factor $Z_F$ contained
in $R_k$.

Next we have to compute the coefficient of the
$\phi\ffs$--term contained in $T_1$ and $T_2$ as a function
of $k$ and as a functional of $R^{(0)}$. This is most easily
done by assuming that $A_\mu^a$ has a
covariantly--constant field strength, i.e., that
$D_\alpha^{ab}F_{\mu\nu}^b=0$, which implies $D_\mu^{ab}F_{\mu\nu}^b=0$,
of course. Though this does not mean that $F_{\mu\nu}^a$ is
$x$--independent, the heat--kernel
$K{(s)}=\exp\left(-s{\cal D}_T\right)$
in such backgrounds is known explicitly \cite{shore}.
Only the first few terms of its expansion in powers of $F_{\mu\nu}$
can contribute to $\ffs$. They read
\be\label{2.19}
\bay
K_{\mu\nu}^{ab}(x,y;s)
& = &
\dis (4\pi s)^{-2}
      \exp\left[-\frac{(x-y)^2}{4s}\right]
\cdot \\
 & \cdot &
\dis \left\{
          \delta_{\mu\nu}\Phi^{ab}(x,y)
          -2i\bar{g}s\Phi^{ac}(x,y)F_{\mu\nu}^{cb}(y)+\dots
          \right\}
\eay
\ee
where
\be\label{2.20}
\Phi(x,y)=P\exp\left[i\bar{g}\int_y^x dz_\mu A_\mu(z)\right]
\ee
is the parallel transport operator along a straight
line from $y$ to $x$ in the adjoint representation. It
satisfies \cite{shore}
\be\label{2.21}
D_\mu^{ab}\Phi^{bc}(x,y)=\frac{i\bar{g}}{2}
\Phi^{ab}(x,y)F_{\mu\nu}^{bc}(y)(x_\nu-y_\nu)
\ee
The actual evaluation of $T_{1,2}$ is somewhat subtle
and one must carefully observe the order of the various
limiting procedures involved. Let $\Omega=\Omega({\cal D}_T)$ be
an arbitrary operator depending on ${\cal D}_T$
with position--space matrix elements $\Omega_{\mu\nu}^{ab}(x,y)$.
We need traces of the type
\be\label{2.22}
\bay
\Tr[{\cal V}\Omega] & \equiv & \dis
\int d^4xd^4y \,{\cal V}_{\mu\nu}^{ab}(x,y) \Omega_{\nu\mu}^{ba}(y,x) \\
&=&\dis
\int d^4x\,\phi(x)\lim_{y\to x}
\Big\{
    \varepsilon_{\alpha\mu\beta\nu}
    D_\alpha^{ca}(x)D_\beta^{cb}(y)
    \Omega_{\mu\nu}^{ab}(x,y)
\\
&&\dis \qquad\qquad\qquad\quad
+ i\bar{g}\,\,
   \mbox{$ F_{\mu\nu}^{ab} \kern -1.65em  \lower -.2ex \hbox{${}^*$}$}
     \kern +1.5em
    (x)
    \Omega_{\mu\nu}^{ab}(x,y)
\Big\}
\eay
\ee
Here we used \rf{2.4} and performed an integration by
parts. In accord with the arguments outlined in the
introduction we dropped the surface term because
the limit $\phi(x)\to1$ is to be performed only at the very
end. Let us assume that $\Omega$ can be represented
as a Laplace transform:
\be\label{2.23}
\Omega({\cal D}_T)=\int_0^\infty ds\, \omega(s) e^{-s{\cal D}_T}
\ee
In our applications this will always be the case and therefore
\be\label{2.24}
\Omega_{\mu\nu}^{ab}(x,y) =\int_0^\infty ds\, \omega(s)
K_{\mu\nu}^{ab}(x,y;s)
\ee
By inserting \rf{2.24} with \rf{2.19} into \rf{2.22} and
making repeated use of \rf{2.21} one finds after
a lengthy calculation
\be\label{2.25}
\Tr[{\cal V}\Omega] = -\frac{\bar{g}^2}{8\pi^2} T(G) L[\omega(s)]
\int d^4x \phi(x)\ffs +\dots
\ee
The functional $L[\omega]$ is defined in terms of a
coincidence limit \mbox{$z\equiv x-y \to 0$ :}
\be\label{2.26}
L[\omega]=\lim_{z\to 0} \frac{z^2}{4} \int_0^\infty
\frac{ds}{s^2}\, \omega(s) \exp\left(-\frac{z^2}{4s}\right)
\ee
We observe that if $\omega(s)$ vanishes sufficiently fast
for $s\to 0$ the integral exists without the exponential
damping factor and we get $L[\omega]=0$ immediately.
The normalization of $L$ is such that $L[\omega]=1$
for a constant function $\omega=1$. Likewise one
obtains for traces involving the projector
$P_L=-(D\otimes D){\cal D}_T^{-1}$:
\be\label{2.27}
\Tr[{\cal V}P_L\Omega] = \frac{\bar{g}^2}{32\pi^2} T(G)
L[\tilde{\omega}(s)/s]
\int d^4x \phi(x)\ffs +\dots
\ee
Since $P_L$ gives rise to an additional factor of
${\cal D}_T^{-1}$ one defines $\tilde{\omega}$ by
${\cal D}_T^{-1}\Omega({\cal D}_T) =
\int_0^\infty ds \, \tilde{\omega}(s) K(s)$.
It is easily expressed in terms of the Laplace
transform of $\Omega$:
\be\label{2.27a}
\tilde{\omega}(s)=s\int_0^1 du\, \omega(su)
\ee
The trace \rf{2.25} is entirely due to the second term
(proportional to $sF_{\mu\nu}^{cd}$) in the curly brackets of
eq.\rf{2.19}. The projected trace \rf{2.27} receives contributions
only from the first term proportional to $\delta_{\mu\nu}$. This
explains the additional factor of $1/s$ in the argument
of $L[\tilde{\omega}(s)/s]$ in \rf{2.27}.

For the computation of $T_1$ it is useful to define
the dimensionless function $\sigma_1$ by
\be\label{2.27b}
\left[y+R^{(0)}(y)\right]^{-1} =\int_0^\infty ds\, \sigma_1(s)
e^{-sy}
\ee
For a momentum independent (mass--type) cutoff\footnote{
  Though this cutoff does not satisfy $R^{(0)}(\infty)=0$ it may be
  used if it does not cause UV divergences \cite{cs}. }
$R^{(0)}(y)=1$, say, it reads $\sigma_1(s)=\exp(-s)$, and
for the exponential cutoff \cite{ah,rw3}
\be\label{2.27c}
R^{(0)}(y) = y \left[e^y-1\right]^{-1}
\ee
it is a step function: $\sigma_1(s)=\th(1-s)$.

After these preparations we are now ready to write down the
relevant term of $T_1$ in \rf{2.18} as a functional of
$\sigma_1(s)$:
\be\label{2.28}
T_1=k^2\left\{
    j_1(k^2)+\frac{1}{4}(1-\alpha)j_1^{(\alpha)}(k^2)
       \right\}
\frac{\bar{g}^2 T(G)}{4\pi^2}\int d^4x\, \phi(x)\ffs
\ee
Here
\be\label{2.29}
j_1(k^2)\equiv -L\left[\frac{d}{dk^2}\sigma_1(k^2s)\right]
\qquad,\qquad
j_1^{(\alpha)}(k^2)\equiv
-L\left[\frac{d}{dk^2}\sigma_1^{(\alpha)}(k^2s)\right]
\ee
with
\be\label{2.30}
\sigma_1^{(\alpha)}(k^2s)\equiv \int_0^1 du\, \sigma_1(k^2su)
\ee
Let us investigate the properties of the function
\be\label{2.31}
j_1(k^2)=-\lim_{z\to0}\frac{z^2}{4}\int_0^\infty
\frac{ds}{s}\exp\left(-\frac{z^2}{4s}\right)\sigma_1^\prime(k^2s)
\ee
(The prime denotes the derivative with respect to the argument.)
In solving the evolution equation \rf{2.17} we shall encounter
integrals of the form
\be\label{2.32}
I=\int_{k_0^2}^\infty dk^2\,j_1(k^2) \varphi(k^2)
\ee
where $k_0>0$ is a constant and $\varphi(k^2)$ is a smooth test
function which does not necessarily vanish at infinity.
In our application the point--separation \mbox{$z\equiv x-y\ne0$}
plays the r\^ole of an UV cutoff. It can be removed
only after the $k^2$--integration has been performed. Hence
\rf{2.32} should be interpreted as
\be\label{2.33}
\bay
I&=&\dis
-\lim_{z\to0}\frac{z^2}{4}
\int_0^\infty \frac{ds}{s}\exp\left(-\frac{z^2}{4s}\right)
\int_{k_0^2}^\infty dk^2\, \varphi(k^2)\sigma_1^\prime(k^2s)
\\
&=&\dis
-\lim_{z\to0}\frac{z^2}{4}
\int_0^\infty \frac{ds}{s}\exp\left(-\frac{1}{s}\right)
\int_{k_0^2}^\infty dk^2\, \varphi(k^2)
\sigma_1^\prime({\text \frac{1}{4}}k^2z^2s)
\eay
\ee
where we rescaled $s\to \frac{1}{4}z^2s$ in the second line.
Setting \mbox{$p^2\equiv \frac{1}{4}k^2z^2s$} one obtains
\be\label{2.34}
I=-\int_0^\infty \frac{ds}{s^2} \exp\left(-\frac{1}{s}\right)
   \lim_{z\to0}
   \int_{\frac{1}{4}z^2k_0^2s}^\infty dp^2\,
   \varphi\left(\frac{4p^2}{z^2s}\right)
   \sigma_1^\prime(p^2)
\ee
Because the $s$--integral is well convergent for both $s\to\infty$
and $s\to0$ it commutes with the limit $z\to0$. For
$\sigma_1$ regular, the $p^2$--integral becomes in this limit
\be\label{2.35}
\varphi(\infty)\int_0^\infty dp^2 \, \sigma_1^\prime(p^2)=
\varphi(\infty)\left[\sigma_1(\infty)-\sigma_1(0)\right]
\ee
Because \rf{2.27b} and $R^{(0)}(0)=1$ imply that
$\int_0^\infty ds \sigma_1(s)=1$ we have $\sigma_1(\infty)=0$.
It is also easy to see that
$\sigma_1(0)=1$
which follows from
\be\label{2.37}
1=\lim_{y\to\infty}\frac{y}{y+R^{(0)}(y)}
 =-\lim_{y\to\infty}\int_0^\infty ds \,
 \sigma_1(s) \frac{d}{ds} e^{-sy}
\ee
after an integration by parts. Inserting \rf{2.35} into \rf{2.34}
leads to the remarkable result that
\be\label{2.38}
\int_{k_0^2}^\infty dk^2\, j_1(k^2)\varphi(k^2)=\varphi(\infty)
\ee
Thus, with the understanding that the
coincidence limit is performed after the integration,
we find that the ``function" $j_1$ actually is a
distribution which has the character of a $\delta$--peak
located at infinity. Though this behavior might seem
strange at first sight it is precisely what one
would expect on physical grounds. As we shall see in
detail later on, the renormalization of the topological
charge by gauge boson loops is a phenomenon which is
very similar to the chiral anomaly of fermions.
In either case the essential physics is contained in
(carefully regularized) short distance singularities of
operator products.

The analysis for $j_1^{(\alpha)}(k^2)$ proceeds along the same
lines with $\sigma_1$ replaced by $\sigma_1^{(\alpha)}$ and
one finds
\be\label{2.39}
\int_{k_0^2}^\infty dk^2 \, j_1^{(\alpha)}(k^2)
\varphi(k^2)=\varphi(\infty)
\ee
One of the interesting properties of the integrals \rf{2.38} and
\rf{2.39} is that they do not depend on the precise form of
the cutoff $R^{(0)}(y)$: they describe {\it universal} properties
of the renormalization group flow. Coming now to the
second piece on the RHS of the evolution equation, $T_2$,
this is not the case any longer. $T_2$ is proportional
to the anomalous dimension $\eta_F$ and contains the higher
order corrections therefore. It is most easily
calculated in terms of the Laplace transform $\sigma_2$
defined by
\be\label{2.40}
R^{(0)}(y)[y+R^{(0)}(y)]^{-2} =\int_0^\infty ds \, \sigma_2(s)
e^{-sy}
\ee
One obtains
\be\label{2.41}
T_2=-k^2\left\{
      j_2(k^2)+\frac{1}{4}(1-\alpha)j_2^{(\alpha)}(k^2)
        \right\}
\eta_F(k)\frac{\bar{g}^2T(G)}{8\pi^2}\int d^4x\, \phi(x)\ffs
\ee
with
\be\label{2.42}
j_2(k^2)\equiv k^{-2}L\left[\sigma_2(k^2s)\right]
\quad , \quad
j_2^{(\alpha)}(k^2)\equiv k^{-2}
L\left[\sigma_2^{(\alpha)}(k^2s)\right]
\ee
and
\be\label{2.43}
\sigma_2^{(\alpha)}(k^2s)\equiv \int_0^1 du\,
\sigma_2(k^2su)
\ee
By an analysis similar to the one above one can derive that
for $0<k_0<\infty$
\be\label{2.44}
\int_{k_0^2}^\infty dk^2\, j_2(k^2)\varphi(k^2)
=
\int_{k_0^2}^\infty dk^2\, j_2^{(\alpha)}(k^2)\varphi(k^2)
=\xi\,\varphi(\infty)
\ee
The constant $\xi$ is given by
\be\label{2.45}
\xi=\int_0^\infty \frac{ds}{s}\sigma_2(s)
   =\int_0^\infty dy\,R^{(0)}(y)[y+R^{(0)}(y)]^{-2}
\ee
We observe that $j_2$ and $j_2^{(\alpha)}$ too are
delta--distributions
with a peak at infinity, but unlike $j_1$ and $j_1^{(\alpha)}$
they are not universal. Their normalization $\xi$
depends on the cutoff function $R^{(0)}$. For $R^{(0)}=1$ one
has $\xi=1$, for instance, and the exponential cutoff \rf{2.27c}
yields $\xi=\ln(2)$. This cutoff or scheme dependence
of the higher order corrections
is a familiar phenomenon \cite{ah}. It cancels at the
level of observable quantities.

Let us now insert $T_1$ and $T_2$ from \rf{2.28} and \rf{2.41} into
the evolution equation. Switching from $k$ to $k^2$ as the evolution
parameter, \rf{2.17} becomes
\be\label{2.46}
\bay
\dis
\frac{d}{dk^2}\th(k) &=&\dis
\th(k) Z_F(k)^{-1}
 \frac{\bar{g}^2}{4\pi^2}T(G)
\Bigg\{
  \left[
     j_1(k^2)+\frac{1}{4}(1-\alpha)j_1^{(\alpha)}(k^2)
  \right]\\
&&\dis \qquad\qquad\qquad\,
  -\frac{1}{2}\eta_F(k)
   \left[
    j_2(k^2)+\frac{1}{4}(1-\alpha)j_2^{(\alpha)}(k^2)
   \right]
\Bigg\}
\eay
\ee
By integrating this equation from an arbitrary $k_0^2>0$ to infinity
and taking advantage of the $\delta$-function nature of the $j$'s,
eqs.\rf{2.38}, \rf{2.39} and \rf{2.44}, one arrives at
\be\label{2.47}
\th(k_0)
=
\left[
  1-\frac{\bar{g}^2}{4\pi^2}T(G)
  \left(1+\frac{1}{4}(1-\alpha)\right)
  \left\{1-\frac{1}{2}\xi\,\eta_F(\infty)\right\}
\right]
\th(\infty)
\ee
This is our final result for all strictly positive
scales $k_0>0$. The $\th$--parameter is renormalized relative
to its bare value $\th(\infty)$ by a finite, $k$--independent factor.
The function $\th(k)$ is constant almost everywhere, but
it has a finite discontinuity at infinity. When compared to
``ordinary" coupling constants such as $g^2(k)$, for instance,
a renormalization group trajectory of this kind is
quite unusual. However, in the appendix we show
in detail that this behavior is precisely the way in which
the pathologies of the triangle anomaly manifest
themselves in the renormalization group framework used here.
The above calculation amounts to computing the
renormalized vacuum expectation value of $\ffs\sim\partial_\mu K_\mu$
where $K_\mu$ denotes the Chern--Simons current.
This calculation has many features in common with
its fermionic counterpart where $K_\mu$ is replaced
by the axial vector current
$J_\mu^5\equiv \bar{\psi}\gamma_\mu\gamma_5\psi$.
The jump of $\th$ and of $<\ffs>$ corresponds to the anomaly
term in $\partial_\mu J_\mu^5$. For a detailed comparison
we refer to \cite{shif,joh,shifcs}. Similar ``bosonic anomalies"
are known  to occur when one quantizes antisymmetric
tensor fields in a gravitational background\cite{ba,phot}.

In the Feynman gauge $\alpha=1$ and with the
higher order corrections neglected $(\eta_F\to0)$, our
eq.\rf{2.47} is consistent with the one--loop results of
refs.\cite{shif} and \cite{joh}. In the truncation used in
this paper we find additional contributions which
partially sum up the effects of the higher loop
orders. They are proportional to
$\eta_F(\infty)\equiv \bar{g}^{-2}\beta_{g^2}(\bar{g})$
with $\beta_{g^2}$ given by \rf{2.16} in terms of the bare
gauge coupling $\bar{g}\equiv g(\infty)$. This suggests that, at
the level of the effective average action, the change of $\th$
is not saturated by its one--loop value.

As for the terms proportional to $(1-\alpha)$, our result \rf{2.47}
coincides neither with \cite{anjoh} nor with \cite{shif}.
These terms originate from the traces of the type $\Tr[P_L(\dots)]$
which describe longitudinal gauge bosons circulating
inside the loops. At first sight the $\alpha$--dependence comes as
a surprise since the background field satisfies $D_\mu F_{\mu\nu}=0$,
i.e., it is ``on shell". However, as a regulating device
we kept $x\ne y$ until the evolution equation
was integrated. In practice a non--zero point--separation
introduces a kind of virtuality similar to a
nonvanishing external momentum square in the case of
the usual diagrammatic calculations based upon
plane waves. Thus the status of the $\alpha$--dependence
is the same as discussed in detail by
Shifman and Vainshtein \cite{shif}.

\mysection{Infrared Renormalization}
Up to now we derived and solved the evolution equation
for $\th(k)$ from infinity down to a scale
$k_0$ which may be chosen arbitrarily low but must
be kept different from zero. It is easy to convince
oneself that the derivation of the previous section does not
hold for the precise equality $k_0=0$. In fact, we are now going
to show that for $k_0\to0$ the function
$\th(k)$ suffers from a second discontinuity \cite{joh}. The physical
origin of this second jump are the zero--modes of the operator
${\cal D}_T$. One of the big advantages of the
method employed here is that the beta--functions of the
generalized couplings ($g$ and $\th$ here) can be determined
by inserting any background field which gives a
nonvanishing value to the relevant field monomials.
The beta--functions do not depend on the background chosen,
and we may use whatever is convenient from a
computational point of view \cite{rw5,ah}. The limit
$k_0\to0$ is most conveniently investigated by inserting
a self--dual field into \rf{2.7} because this will
allow us to recast the problem in a fermionic language
and powerful index theorems become available.
Because $F_{\mu\nu}=\fsn_{\mu\nu}$ implies $D_\mu F_{\mu\nu}=0$,
the simplifications of the evolution equation made in sect.2 are still
allowed, and we can rewrite \rf{2.7} as
\be\label{3.1}
\begin{array}{l}
\dis
\frac{d}{dk^2}\th(k)\int d^4x\,\phi(x)\ffs
=
\\
\dis
\qquad\qquad
-2\th(k)\Tr
\left[
     {\cal V}\Big(Z_F(k){\cal D}_T +R_k(\Delta)\Big)^{-2}
     \frac{d}{dk^2}R_k(\Delta)
\right]
\eay
\ee
For simplicity we set $\alpha=1$ in this section, i.e.,
${\cal D}={\cal D}_T$.
It is obvious from \rf{3.1} that a zero eigenvalue of
${\cal D}_T$ produces a highly divergent contribution to
the trace when \mbox{$R_k\sim k^2\to 0$}. We shall see that this
leads to the discontinuity of $\th(k)$ mentioned above. While it
is true that a field satisfying $F_{\mu\nu}=\fsn_{\mu\nu}$ cannot
disentangle the invariants $F_{\mu\nu}F_{\mu\nu}$ and
$F_{\mu\nu}\fsn_{\mu\nu}$
the function $Z_F(k)$ is continuous for $k\to 0$ and hence any
nontrivial behavior for $k\to0$ should be attributed to $\th(k)$.

For self--dual backgrounds the technology of refs.\cite{dv}
and \cite{joh} simplifies the analysis, and we start by
defining four $4\times4$--matrices $\Omega_\mu$
\renewcommand{\arraystretch}{1.2}
\be\label{3.2}
\left(\Omega_\mu\right)_{\alpha\beta}
=
\left\{
\begin{array}{rcl}
\dis
\eta_{\alpha\beta\mu} & \dis {\rm if} &\dis \alpha=1,2,3 \\
-\delta_{\beta\mu}    & \dis {\rm if} &\dis \alpha=4
\end{array}
\right.
\ee
where $\eta_{\alpha\,\mu\nu}(\alpha=1,2,3\,;\,\mu,\nu=1,\dots,4)$
is 't Hooft's symbol \cite{th}. Using its well--known properties one
can derive that
\renewcommand{\arraystretch}{2.0}
\be\label{3.3}
\bay
\dis
\left(\Omega_\mu\Omega_\nu^T\right)_{\alpha\beta}
&\dis =
&\dis
\delta_{\mu\nu}\delta_{\alpha\beta}
+\delta_{\alpha\mu}\delta_{\beta\nu}
-\delta_{\alpha\nu}\delta_{\beta\mu}
+\varepsilon_{\mu\nu\alpha\beta}
\\
\dis
\left(\Omega_\mu^T\Omega_\nu\right)_{\alpha\beta}
&\dis =
&\dis
\delta_{\mu\nu}\delta_{\alpha\beta}
+\delta_{\alpha\mu}\delta_{\beta\nu}
-\delta_{\alpha\nu}\delta_{\beta\mu}
-\varepsilon_{\mu\nu\alpha\beta}
\eay
\ee
If we set $\hat{D}\equiv \Omega_\mu D_\mu$,
$\hat{D}^T\equiv \Omega_\mu^T D_\mu$ and use the condition
$F_{\mu\nu}=\fsn_{\mu\nu}$ then \rf{3.3} implies
\be\label{3.4}
\bay\dis
\left(\hat{D}^T \hat{D}\right)_{\alpha\beta}
&=& \dis D^2 \delta_{\alpha\beta}
\\
\dis
\left(\hat{D} \hat{D}^T\right)_{\alpha\beta}
&=& \dis D^2 \delta_{\alpha\beta}-2i\bar{g}F_{\alpha\beta}
\eay
\ee
Thus ${\cal D}_T=-\hat{D} \hat{D}^T$ for self--dual fields. Using the
$\Omega_\mu$'s as building blocks, we introduce the following
$8\times8$--matrices:
\renewcommand{\arraystretch}{1.2}
\be\label{3.5}
\Gamma_\mu =
\left[
  \begin{array}{cc}
   \dis 0 & \dis \Omega_\mu \\
   \dis -\Omega_\mu^T & \dis 0
  \eay
\right]
\ee
By virtue of the relations
\renewcommand{\arraystretch}{2.0}
\be\label{3.6}
\bay
\dis
\Omega_\mu\Omega_\nu^T +\Omega_\nu\Omega_\mu^T
&=&\dis 2\delta_{\mu\nu}
\\
\dis
\Omega_\mu^T\Omega_\nu +\Omega_\nu^T\Omega_\mu
&=&\dis 2\delta_{\mu\nu}
\eay
\ee
the $\Gamma_\mu$'s are seen to constitute an 8--dimensional
representation of the 4--dimensional Clifford algebra:
\be\label{3.7}
\Gamma_\mu\Gamma_\nu+\Gamma_\nu\Gamma_\mu
=-2\delta_{\mu\nu}
\ee
Note that $\Gamma_\mu^\dagger=-\Gamma_\mu$ because $\Gamma_\mu$
is real and antisymmetric. An important r\^ole will be played by
the ``chirality" operator
\renewcommand{\arraystretch}{1.0}
\be\label{3.8}
\Gamma_5\equiv -\Gamma_1\Gamma_2\Gamma_3\Gamma_4
=
\left[
 \begin{array}{cc}
  \dis 1 &\dis\makebox[15pt]{ 0} \\
  \dis 0 &\dis\makebox[15pt]{ -1}
 \eay
\right]
\ee
It has the usual properties $\Gamma_5^2=1$ and
$\{\Gamma_5,\Gamma_\mu\}=0$.
In order to reformulate the evolution equation in a
``fermionic" language we need the Dirac operator
\renewcommand{\arraystretch}{1.2}
\be\label{3.9}
\DD = \Gamma_\mu D_\mu =
\left[
 \begin{array}{cc}
  \dis\makebox[20pt]{0} & \dis\makebox[20pt]{$\hat{D}$} \\
  \dis\makebox[20pt]{$-\hat{D}^T$} & \dis\makebox[20pt]{0}
 \eay
\right]
\ee
Because $\DD^2$ is a  block--diagonal matrix, we find the
useful relation
\be\label{3.10}
\tr_8
\left[
   \Gamma_5\Gamma_\mu\DD\, G(\DD^2)
\right]
=
-\tr_4
\left[
  \Omega_\mu\hat{D}^T \,G(-\hat{D}\hat{D}^T)
 -\Omega_\mu^T\hat{D}\, G(-\hat{D}^T\hat{D})
\right]
\ee
Here $G$ is an arbitrary function and $\tr_4$ and $\tr_8$ denote
the traces with respect to the $4\times4$ and the $8\times8$ matrix
structures, respectively. From the identity
\be\label{3.11}
\left(
  \Omega_\mu\Omega_\nu^T
 -\Omega_\mu^T\Omega_\nu
\right)_{\alpha\beta}
=
2\varepsilon_{\mu\nu\alpha\beta}
\ee
one obtains the ``spinor" representation of the operator ${\cal V}$:
\be\label{3.12}
{\cal V}_{\alpha\beta} =
\frac{1}{2}\left(\partial_\mu \phi\right)
\left[
 \Omega_\mu \hat{D}^T-\Omega_\mu^T \hat{D}
\right]_{\alpha\beta}
\ee
The evolution equation contains a trace of the form
\renewcommand{\arraystretch}{2.0}
\be\label{3.13}
\kern-.5em
\begin{array}{l}
\dis
\Tr[{\cal V}G({\cal D}_T)]
=
-\frac{1}{2}\int d^4x\, \phi(x) \tr_c \tr_4 \partial_\mu
\!<x\!\vert
\left\{\Omega_\mu\hat{D}^T-\Omega_\mu^T\hat{D}\right\}
G(-\hat{D}\hat{D}^T)\vert x\!>
\\
\dis
=
-\frac{1}{2}\int d^4x\, \phi(x) \tr_c \tr_4 \partial_\mu
\!<\!x\vert
 \Omega_\mu  \hat{D}^T G(-\hat{D}  \hat{D}^T)
-\Omega_\mu^T\hat{D}   G(-\hat{D}^T\hat{D}  )
\vert x\!>-\Delta T
\eay
\ee
($\tr_c$ denotes the trace in color space.)
As in sect.2 we performed an integration by parts and
assumed that $\phi(x)$ falls off sufficiently fast so that
there are no surface terms. In the last line of \rf{3.13} we
added and subtracted the same terms, i.e., $\Delta T$ is given
by \cite{joh}
\be\label{3.14}
\Delta T =
\frac{1}{2}\int d^4x\, \phi(x) \tr_c \tr_4 \partial_\mu
\!<\!x\vert
 \Omega_\mu^T\hat{D}
\left\{
G(-\hat{D}^T\hat{D}  ) - G(-\hat{D}  \hat{D}^T)
\right\}
\vert x\!>
\ee
Provided $G$ is chosen in such a way that the trace actually
exists one can use the method of section 2 together with
the selfduality condition to show that $\Delta T$ does not
contribute to the $\phi F_{\mu\nu}\fsn_{\mu\nu}$--term and can be
neglected therefore. The remaining terms in \rf{3.13} have
the structure of \rf{3.10}. Hence the whole trace
can be rewritten in the language of 8--component
spinor matrices:
\be\label{3.15}
\begin{array}{rcl}
\dis
\Tr[{\cal V}G({\cal D}_T)]
&=&\dis
\frac{1}{2}\int d^4x\, \phi(x)\, \tr_c \tr_8 \partial_\mu
\!<\!x\vert
\Gamma_5\,\Gamma_\mu\,\DD\, G(\DD^2)
\vert x\!>
\\
&=&\dis
\int d^4x\, \phi(x)\, \tr_c \tr_8
\!<\!x\vert
\Gamma_5\,\DD^2\, G(\DD^2)
\vert x\!>
\eay
\ee
By using \rf{3.15} in \rf{3.1} we arrive at the desired representation
of the evolution equation:
\be\label{3.16}
\kern-.7em
\begin{array}{l}
\dis
\frac{d}{dk^2}\th(k)F_{\mu\nu}^a(x)\fs(x)
=
\dis
-2\th(k)\tr_c\tr_8
<\!x\vert
\Gamma_5\DD^2\!
\left[
  Z_F(k)\DD^2+Z_F(k)k^2 R^{(0)}(\DD^2\! / k^2)
\right]^{-2}
\\
\dis
\qquad\qquad
\qquad\qquad
\qquad\qquad
\qquad\qquad
\cdot
\frac{d}{dk^2}
\left\{
  Z_F(k)k^2 R^{(0)} (\DD^2\!/k^2)
\right\}
\vert x\!>
\eay
\ee
Let us pause here for a moment and recall
the Atiyah--Singer index theorem for the operator
$\DD$ \cite{shind,joh}. We assume that spacetime is a large
4--sphere.
Hence the spectrum is discrete and for a given background $A_\mu$
there are $n_+[A]\,(n_-[A])$ zero modes $\psi_+\,(\psi_-)$ of
chirality $+1\, (-1)$.
We adopt the usual definitions $\psi_\pm =P_\pm\psi_\pm$
with the projectors $P_\pm=\frac{1}{2}\left(1\pm\Gamma_5\right)$.
One has for all $t>0$
\be\label{3.17}
n_+[A]-n_-[A]=\Tr\left[\Gamma_5\exp\left(-t\DD^2\right)\right]
\ee
because by a standard argument \cite{as} the non--zero modes
cancel in the trace. By inserting the heat--kernel expansion
for $\DD^2$ and letting $t\to0$ one easily arrives at the
index theorem
\be\label{3.18}
\bay
\dis
n_+[A]-n_-[A]&=&\dis 4 T(G) Q \\
&=&\dis
T(G)\frac{\bar{g}^2}{8\pi^2}\int d^4x\,\ffs
\eay
\ee
The prefactor $4T(G)$ of the topological charge arises since
we are dealing with ``fermions" in the adjoint representation
and because we employ
a non--standard representation of the Clifford algebra.\footnote{
It enters the heat--kernel computation of the index via the
identity
\linebreak
$\tr_8[\Gamma_5\Gamma_\mu\Gamma_\nu\Gamma_\rho\Gamma_\sigma]
=-8\varepsilon_{\mu\nu\rho\sigma}$.
}
The solutions to the zero--mode equation $\DD \psi_\pm=0$ have
the form $\psi_+=(\phi_+,0)$ and $\psi_-=(0,\phi_-)$ where
$\phi_+$ and $\phi_-$ satisfy $\hat{D}^T\phi_+=0$ and
$\hat{D}\phi_-=0$, respectively.
Multiplying by $\hat{D}$ and $\hat{D}^T$ from the left we see that
${\cal D}_T\phi_+=0$ and $D^2\phi_-=0$. Because $D^2$ has no
zero--modes one has $n_-=0$ so that $n_+=4T(G)Q>0$
is the number of zero--modes of ${\cal D}_T$.\footnote{
There are no solutions of ${\cal D}_T\phi_+=0$ with
$\hat{D}^T\phi_+\neq0$ because there exists
a positive--definite inner product with respect to which
$\hat{D}^T$ is the adjoint of $-\hat{D}$.
}

Equipped with the index theorem it is easy to analyze
the evolution equation \rf{3.16}. Since we already know
from section 2 that $\th(k)$ is constant even for $k$
close to (but different from) zero, it is sufficient
to integrate \rf{3.16} from zero up to a small $k_0^2>0$.
Because $Z_F(k)$ is continuous for $k\to0$ we may
replace $Z_F(k)$ by $Z_F(0)$ in \rf{3.16}. Thus
\be\label{3.19}
[\th(k_0)-\th(0)]F_{\mu\nu}(x)\fs(x) =
-2<x\vert \tr_c\tr_8 \Gamma_5 Y(\DD^2)\vert x>
\ee
with
\be\label{3.20}
Y(\lambda)\equiv -\lambda Z_F(0)^{-1}
\int_0^{k_0^2} dk^2 \, \th(k)
\frac{d}{dk^2}
\left[\lambda+k^2R^{(0)}(\lambda/k^2)\right]^{-1}
\ee
Seen as a function of the real parameter $\lambda$, $Y$
has a smooth limit for $\lambda \to0$. Writing
\be\label{3.21}
Y(\lambda)=-Z_F(0)^{-1}
\int_0^{k_0^2/\lambda} dy\,
\th\left(\lambda^\frac{1}{2} y^{\frac{1}{2}}\right)
\frac{d}{dy}\left[1+yR^{(0)}(1/y)\right]^{-1}
\ee
we observe that the constant factor $\th(0)$ emerges in the limit
$\lambda\to 0$, and that $y$ is integrated from zero to infinity. One
obtains the $R^{(0)}$--independent limit
\be\label{3.22}
Y(0)=Z_F(0)^{-1}\th(0)
\ee
We determine the discontinuity $\th(k_0)-\th(0)$ by
integrating \rf{3.19} over $x$ and applying the index theorem.
The RHS of \rf{3.19} becomes $-2\Tr[\Gamma_5 Y(\DD^2)]$,
and because the non--zero modes of $\DD^2$ with positive
and negative chirality are always paired this equals
$-2Y(0)\Tr[\Gamma_5]$. A regularized version of $\Tr[\Gamma_5]$
is provided by
\rf{3.17} so that we may replace $\Tr[\Gamma_5]$ by $4T(G)Q$.
Putting everything together we arrive at
the final answer for the jump of $\th(k)$ near $k=0$:
\be\label{3.23}
\th(0)=
\left[
  1-\frac{1}{4\pi^2}T(G)\bar{g}^2Z_F(0)^{-1}
\right]^{-1}\th(k_0)
\ee
Recall that $\bar{g}^2Z_F(0)^{-1}=g^2(0)$ is the running gauge
coupling at zero momentum and $\bar{g}^2\equiv g^2(\infty)$ is the
bare one. We can combine \rf{3.23} with \rf{2.47} for $\alpha=1$
and express the renormalized $\th$--parameter $\th(0)$ in terms
of the bare parameter $\th(\infty)$:

\be\label{3.24}
\th(0)=
\left[1-\frac{T(G)}{4\pi^2}g^2(0)\right]^{-1}
\left[
  1-\frac{T(G)}{4\pi^2}g^2(\infty)
  \left\{1-\frac{1}{2}\xi\,\eta_F(\infty)\right\}
\right]
\th(\infty)
\ee
This is our main result. It shows that if one understands
the ``renormalization of the $\th$--parameter" in the sense
of performing the limit $\phi(x)\to1$ {\it after}
the theory has been quantized, i.e., after the evolution
equation has been solved, then there is indeed a (finite)
difference between the bare and the renormalized
$\th$--parameter.

The use of an exact renormalization group equation
with the truncation \rf{2.1} amounts to a renormalization
group improved one--loop calculation. Let us switch off
for a moment the corrections which go beyond a standard
one--loop calculation of $\th(0)$. In this case there is no running
of $Z_F$, i.e., $g^2(0)=g^2(\infty)$, and the term in \rf{3.24}
proportional to $\eta_F$ is absent. We see that in this
case $\th(0)=\th(\infty)$ because the discontinuities at
$k=\infty$ and at $k=0$ cancel precisely and there is no net
effect left. This is the compensation which was found in
ref.\cite{joh} by different methods. It can be understood
in close analogy with the well--known argument
which relates the Atiyah--Singer index theorem to the
anomaly equation
\be\label{3.25}
\partial_\mu<\bar{\psi}\gamma_\mu\gamma_5\psi>
=2m<\bar{\psi}\gamma_5\psi>
 +\frac{\bar{g}^2}{16\pi^2}\ffs
\ee
If one integrates this equation over a compact spacetime the
LHS vanishes and the second term on the RHS yields
twice the topological charge. The first term on the RHS
equals $-2\Tr[\gamma_5m(\DD+m)^{-1}]$ and becomes $-2(n_+-n_-)$
in the limit $m\to 0$. Hence $(n_+-n_-)-Q=0$. From our
discussion of fermions in the appendix it is clear that
the compensation of the jumps at $k=0$ and $k=\infty$ is
completely analogous to the compensation of $(n_+-n_-)$
and $Q$. The piece $(n_+-n_-)$ coming from the ``soft"
operator corresponds to the jump at $k\to 0$ and the
``hard" contribution from $F_{\mu\nu}\fsn_{\mu\nu}$ is related
to the jump at infinity.

The calculation in this paper goes beyond a
one--loop computation in that it retains the
running of a second coupling, $g^2(k)$. At this level
of accuracy we find clear evidence for a
nonvanishing renormalization of the $\th$--parameter.
Though the non--trivial running occurs only
in the extreme ultraviolet and infrared,
the discontinuities of $\th(k)$ triggered there do not
compensate any longer.

\mysection{Conclusion}
In this paper we considered the topological term $\int d^4x \ffs$
as the limit of the non--topological interaction  $\int d^4x \phi(x) \ffs$.
We saw that if the limit $\phi(x)\to1$ is taken {\it before} the
renormalization group evolution, or in other words, before the
quantization, then the $\th$--parameter is not renormalized.
This is in accord with the expectations based upon the
interpretation of $\th$ as the quasi--momentum related to
the ``Bloch waves" of the Yang--Mills vacuum. We have also seen
that if one performs the limit {\it after} the evolution
then a nontrivial renormalization of $\th$ occurs. We were
mostly concerned with pure Yang--Mills theory where $\th$ is renormalized
multiplicatively by a finite factor. If one adds fermions (see the appendix)
then there is an additional finite shift of $\th$. We investigated the
beta--function which describes the running of $\th(k)$. Nontrivial
effects are confined to the extreme ultraviolet region where the
anomaly of the Chern--Simons current gives rise to a finite discontinuity,
and to the extreme infrared where the zero modes of the inverse gauge
boson propagator trigger another discontinuity. At the one--loop level
the two discontinuities cancel. In our more refined
calculation which keeps track of the running of both $\th(k)$ and
$g(k)$, the cancellation is incomplete and the renormalized quantity
$\th(0)$ differs from the bare value $\th(\infty)$. The basic mechanism
which spoils the compensation is that the two jumps of $\th(k)$ occur
at very different scales and involve the running gauge coupling
$g(k)$ at different scales therefore.

It is one of the virtues of our renormalization group approach that it
allows for a clear separation of these two regimes.
This is particularly important if one thinks of realistic applications
to QCD, for instance. The running of  $\th$ in the UV can be
reliably calculated with truncations such as the one used in this paper.
The derivation of the discontinuity at $k=0$ rests on much less solid
ground. As a first step to extend the validity towards the
infrared, one could use the more general truncations on which
our investigation of the gluon condensation \cite{rw5} was based.

The discontinuous evolution of $\th(k)$ is closely related to
a similar phenomenon in pure 3--dimensional Chern--Simons
theory. In ref.\cite{cs} we showed that the well--known shift of the
Chern--Simons parameter \cite{pisa} is also due to a renormalization
group trajectory with a discontinuity at $k=0$. In view of the
discussion in ref.\cite{shifcs} this similarity is quite natural.

Remaining is the question of what is the ``correct" way of treating
the topological term. Is the limit $\phi(x)\to1$ to be taken before
or after the evolution? The answer is that it depends on the
physical situation. If the term $S_{\rm top}$ of \rf{1.0} is part of
the bare action then there is certainly no reason to artificially
introduce the $\phi$--field, and $\th$ is not renormalized
therefore. If, however, the topological term arises from an
interaction term $\phi(x)\ffs$ because some pseudoscalar $\phi(x)$
acquires an $x$--independent vacuum expectation value, then the second
alternative applies and $\th$ can be renormalized.

It is quite tempting to speculate that renormalization
effects of $\th$ might provide a solution to the strong
CP--problem, i.e., that they explain why the $\th$--angle
observed in nature is extremely small or zero
while a value of order unity would seem much more
natural. In such a scenario one would have to show
that for any bare parameter $\th(\infty)$ the renormalization
group trajectory ends at a $\th(0)$ which is (close to) zero.
Our results suggest that the effect of the gluons
is much more important than that of the quarks
in this respect. If we assume that they are all massive,
they shift $\th(\infty)$ in the ultraviolet, but they play
no role in the infrared. Also the UV--effects by the gluons are
of a perturbative nature and not very important probably.
However, the zero modes of the inverse gluon propagator
could have a significant impact on $\th(0)$. They act
in the strong coupling regime at a large value of $g^2$.
In fact, if we naively set $g^2(0)=\infty$ in \rf{3.24}
we find that $\th(0)=0$ for all bare parameters $\th(\infty)$!
Clearly it is premature to take this result too seriously
since the truncation we used is by far too simple to allow for
a realistic description of QCD at small momenta.
Nevertheless our result indicates that such a scenario is
possible in principle, and that it is worthwhile to study
this mechanism with improved approximations.
It is interesting that recent lattice investigations
\cite{lat} and low dimensional toy models
\cite{lms} also seem to support the idea that the strong
CP problem could be solved within the standard model itself.

\vspace{8mm}         \noindent
Acknowledgement: I would like to thank W.Dittrich, J.L\"offelholz,
A.Morozov, H.Osborn, J.Pawlowski, G.Schierholz, C.Wetterich and
T.Ziegenhagen for helpful discussions.

   \begin{appendix}

\section*{Appendix}
\newcommand{\sect}[1]{\setcounter{equation}{0}
  \renewcommand{\theequation}{#1.\arabic{equation}}
 }
\sect{A}
In this appendix we study the influence of fermions on
the renormalization of the topological charge. We couple
one flavor of Dirac fermions (in the fundamental representation
of $G$) to $A_\mu^a$, and we assume that
there is also a coupling of the fermions to the external
field $\phi(x)$. We investigate both a pseudovector coupling
$(\partial_\mu\phi)\bar{\psi}\gamma_\mu\gamma_5\psi$
and a pseudoscalar coupling $\phi\bar{\psi}\gamma_5\psi$.
The pseudovector case is the fermionic analogue of
the purely bosonic effects studied in the main body
of this paper. It is closely related to the standard chiral
anomaly, but we include its discussion here because it
is quite interesting to see its similarities and differences
to the "bosonic anomaly" of section 2. The pseudoscalar
coupling, on the other hand, has strikingly different
properties: it leads to a smooth running of $\th(k)$ at all
scales $k$.

Starting with the pseudovector coupling we generalize
the truncation by adding the following term to the $\Gamma_k$ of
eq.\rf{2.1}:
\be\label{a.1}
\Delta\Gamma_k[A,\psi,\bar{\psi};\phi] =
\int d^4x
\left\{
  Z_\psi(k)\, \bar{\psi}\left[\DD+m(k)\right]\psi
 -i\partial_\mu\, \phi(x) \bar{\psi} \gamma_\mu\gamma_5\psi
\right\}
\ee
We determine the scale dependence of the induced
interaction $\sim\phi F_{\mu\nu}\fsn_{\mu\nu}$ by
solving the evolution
equation for the coupled gauge field/fermion system.
Its general form can be found in ref.\cite{wet}. Here the
situation simplifies because in order to determine $\th(k)$
and $Z_F(k)$ backgrounds of the type $A,\bar{A}\neq0$,
$\psi=0=\bar{\psi}$ are sufficient. In this case the
RHS of the evolution equation is simply the sum of the two
traces which are present in \rf{1.1} plus a
similar term involving the fermion--fermion
submatrix of $(\Gamma_k+\Delta\Gamma_k)^{(2)}$. Hence the
running of $\th$ is governed by
\be\label{a.2}
\begin{array}{l}
\dis
i\frac{\bar{g}^2}{32\pi^2}k\frac{d}{dk}\th(k)
\int d^4x\,\phi(x)\ffs
= \\
\dis
\qquad\qquad\qquad\qquad
-\Tr\left\{
    \left[
     \left(\Gamma_k+\Delta\Gamma_k\right)_{\bar{\psi}\psi}^{(2)}
      +R_k
    \right]^{-1}
   k\frac{d}{dk}R_k
    \right\}+\dots
\eay
\ee
The dots represent the contributions of the gauge field
and of the ghosts which we have evaluated already.
A cutoff appropriate for Dirac fermions is
\be\label{a.3}
R_k=Z_\psi(k)kR^{(0)}\left(-\DD^2/k^2\right)
\ee
Expanding the RHS of \rf{a.2} to first order in $\phi$ one
obtains
\be\label{a.4}
\kern-0.8em
\begin{array}{rcl}
\dis
\frac{\bar{g}^2}{32\pi^2}\frac{d}{dk^2}\th(k)
F_{\mu\nu}^a(x)\fs (x)
&=&
\dis
-2Z_\psi(k)^{-1}\tr \gamma_5
 <\!x\vert \DD\left[\DD+\mu_k(\DD^2)\right]^{-2}
\\
&&
\dis
\qquad\qquad\qquad
\cdot
\frac{d}{dk^2}\left[kR^{(0)}(-\DD^2/k^2)\right]\vert x\!>
+{\cal O}(\eta_\psi)
\\
&=&
\dis
Z_\psi(k)^{-1}\frac{d}{dk^2}
\tr \gamma_\mu\gamma_5\gamma_\nu
\lim_{y\to x}
\left[D_\mu(x)+D_\mu^\dagger(y)\right]D_\nu(x)
\\
&&
\dis
\cdot
<x\!\vert \left(-\DD^2+\mu_k(\DD^2)^2\right)^{-1}\vert y \!>
+{\cal O}(\eta_\psi)
+{\cal O}(\partial_k m)
\end{array}
\ee
with $\mu_k(\DD^2)\equiv m+k R^{(0)}(-\DD^2/k^2)$. Here we are
interested in the main features only and do not evaluate
the higher order terms proportional to
$\eta_\psi=-k\frac{d}{dk} \ln Z_\psi$.
In the last line of \rf{a.4} we also neglected the $k$--dependence
of $m$. Since mass effects play no import r\^ole we set
$m(k)=m=$const from now on. For dimensionless variables
$y$ and $\kappa$ we introduce the Laplace transform $\sigma_\psi$ by
\be\label{a.5}
\left[y+\left(\kappa+R^{(0)}(y)\right)^2\right]^{-1}
=
\int_0^\infty ds\, \sigma_\psi(s;\kappa) e^{-sy}
\ee
It satisfies $\sigma_\psi(0;\kappa)=1$ and
$\sigma_\psi(\infty;\kappa)=0$. The operator
appearing in \rf{a.4} can be expressed in terms of
$K(s)\equiv \exp\left(s\DD^2\right)$:
\be\label{a.6}
\left(-\DD^2+\mu_k(\DD^2)^2\right)^{-1}
=\int_0^\infty ds\, \sigma_\psi(sk^2; m/k) K(s)
\ee
The heat--kernel for covariantly constant
fields is well--known \cite{js}. The terms relevant in
the present context are
\be\label{a.7}
\begin{array}{l}
\dis
K(x,y;s)=(4\pi s)^{-2}
\exp\left[-\frac{(x-y)^2}{4s}\right]\Phi(x,y)
\\
\dis
\qquad\qquad\qquad
\cdot
\left\{
   1-\frac{1}{2}i\bar{g}s F_{\mu\nu}(y)\gamma_\mu\gamma_\nu
   -\frac{1}{8}\bar{g}^2s^2
    \left(F_{\mu\nu}(y)\gamma_\mu\gamma_\nu\right)^2+\dots
\right\}
\eay
\ee
Note that the parallel transport operator $\Phi(x,y)$ and
$F_{\mu\nu}(y)$ are matrices in the fundamental representation.
Using \rf{a.6} with \rf{a.7} in \rf{a.4} one finds after
some calculation
\be\label{a.8}
\frac{d}{dk^2}\th(k)=2 Z_\psi(k)^{-1}j_\psi(k^2)
\ee
with
\be\label{a.9}
j_\psi(k^2)\equiv -L
\left[\frac{d}{dk^2}\sigma_\psi(sk^2;m/k)\right]
\ee
By a reasoning similar to the one following eq.\rf{2.31} one
can show that
\be\label{a.10}
\int_{k_0^2}^\infty dk^2\, j_\psi(k^2) \varphi(k^2)
=\varphi(\infty)
\ee
Thus we find the same phenomenon as in the gauge field case:
$j_\psi$ is a $R^{(0)}$--independent $\delta$--peak at infinity.
Moreover, in \rf{a.10} all dependence on the physical fermion mass
$m$ has disappeared.
From\rf{a.8} with \rf{a.10} we get the following result for the
renormalization of $\th$ by the fermions alone $(k_0>0)$:
\be\label{a.11}
\th(k_0)=\th(\infty)-2Z_\psi(\infty)^{-1}
\ee
(Usually one sets $Z_\psi(\infty)=1$.) On the RHS of \rf{a.8} one
should add the contribution from the gauge bosons which was found
in section 2. Doing this, $\th(\infty)$ in \rf{a.11} becomes
multiplied by the square bracket in \rf{2.47}.

Next we look at the impact the zero--modes of $\DD$ have
on $\th(k)$. As \rf{a.11} is valid for $k_0$ arbitrarily
close to zero, at most they can lead to a discontinuity at $k=0$.
We integrate \rf{a.4} from $0$ to a nearby point $k_0$:
\be\label{a.12}
\begin{array}{l}
\dis
\frac{\bar{g}^2}{32\pi^2}
\left[\th(k_0)-\th(0)\right]
\int d^4x \ffs
\\
\dis
\qquad
\quad
=2\int_0^{k_0^2} dk^2 \, Z_\psi(k)^{-1}
\frac{d}{dk^2}
\Tr\left[
      \gamma_5\DD
      \left(\DD+m+kR^{(0)}(-\DD^2/k^2)\right)^{-1}
   \right]
\\
\dis
\qquad
\quad
=-2Z_\psi(0)^{-1}\Tr\left[\gamma_5\frac{\DD}{\DD+m}\right]
\eay
\ee
Because the chiralities of its excited states are always
paired, only the zero modes of $\DD$ contribute to the last
trace in \rf{a.12}. Hence for $m\neq0$ this trace vanishes
and for $m=0$ its value is given by the index
theorem \mbox{$\Tr[\gamma_5]\equiv n_+-n_-=Q$}. Therefore
we obtain for the behavior near $k=0$
\renewcommand{\arraystretch}{1.2}
\be\label{a.13}
\th(0)=
\left\{
 \begin{array}{lcc}
   \dis \th(k_0)                 & \dis {\rm if}&\dis m\neq 0\\
   \dis \th(k_0)+2Z_\psi(0)^{-1} & \dis {\rm if}&\dis m= 0
  \end{array}
\right.
\ee
We see that for the massless fermions
\mbox{$\th(0)=\th(\infty)+2[Z_\psi(0)^{-1}-Z_\psi(\infty)^{-1}]$}.
Only if one neglects the running of $Z_\psi$ the two discontinuities
cancel. For a Dirac fermion there is no general reason for
$m(k=0)$ to vanish , and contrary to the situation with the gauge boson
the jump at $k=0$ is more the exception than the rule.
If one includes the gauge boson contribution in \rf{a.12} the RHS of
\rf{a.13} contains an additional factor of
\mbox{$[1-g^2(0)T(g)/4\pi^2]^{-1}$}.

Finally let us see what happens if we introduce a ``soft"
coupling of $\phi(x)$ to the pseudoscalar $\bar{\psi}\gamma_5\psi$.
We replace \rf{a.1} by
\be\label{a.14}
\Delta\Gamma_k\left[A,\psi,\bar{\psi};\phi\right]
=
\int d^4x\,
\left\{
  Z_\psi(k)\bar{\psi}[\DD+m(k)]\psi
 +im_5(k)\phi(x)\bar{\psi}\gamma_5\psi
\right\}
\ee
where $m_5$ is a (possibly $k$--dependent) coupling with
the dimension of a mass. Proceeding as above and defining
$\omega_5(s)$ by

\be\label{a.15}
\frac{m+kR^{(0)}(-\DD^2/k^2)}
     {-\DD^2+\left[m+kR^{(0)}(-\DD^2/k^2)\right]^2}
=
\int_0^\infty ds\, \omega_5(s) K(s)
\ee
one obtains (up to terms proportional to $\partial_k m$)
\renewcommand{\arraystretch}{2.0}
\be\label{a.16}
\frac{\bar{g}}{32\pi^2}\frac{d}{dk}\th(k)\ffs
=
-m_5(k) Z_\psi(k)^{-1}
\frac{d}{dk}\int_0^\infty ds \, \omega_5(s)
<x\!\vert \tr \gamma_5 K(s) \vert x\!>
\ee
Upon inserting the diagonal matrix element of the
heat--kernel \rf{a.7} this leads to
\be\label{a.17}
\bay
\dis
\frac{d}{dk}\th(k)
& = &\dis
-m_5(k)Z_\psi(k)^{-1}\frac{d}{dk}
\int_0^\infty ds\, \omega_5(s)
\\
&=&\dis
- m_5(k)Z_\psi(k)^{-1}\frac{d}{dk}(m+k)^{-1}
\eay
\ee
The second line of \rf{a.17} obtains be setting $\DD\to0$ in
\rf{a.15} and using $R^{(0)}(0)=1$. This time we find
a smooth evolution of $\th$ which is governed by the
equation (leaving gauge field effects aside)
\be\label{a.18}
\frac{d}{dk}\th(k)=
\frac{m_5(k)}{Z_\psi(k)[m+k]^2}
\ee
It is remarkable that also this evolution is universal, i.e.,
independent of the shape of $R^{(0)}$. Eq.\rf{a.18} is trivial
to solve if we approximate $Z_\psi(k)\equiv 1$ and $m_5(k)\equiv m_5$:
\be\label{a.19}
\th(k)=\th(\infty)-\frac{m_5}{m+k}
\ee
For massless fermions there is (at least within this approximation)
a singularity $\sim 1/k$ as $k$ approaches zero.
For $m\neq0$ the limit is finite:
\be\label{a.20}
\th(0)=\th(\infty)-\frac{m_5}{m}
\ee
There exists a distinguished value for the coupling
$m_5$, namely $m_5=2m$. For this value, the pseudovector
coupling term in \rf{a.1} is related to its pseudoscalar
counterpart in \rf{a.14} by the classical divergence equation
\mbox{$\partial_\mu(\bar{\psi}\gamma_\mu\gamma_5\psi)
       =2m(\bar{\psi}\gamma_5\psi)$}. From \rf{a.20} we get
\mbox{$\th(0)=\th(\infty)-2$} in this case, but exactly the same
relation also follows from \rf{a.11} for $Z_\psi=1$ and $k_0=0$.
(Recall that there is no jump at $k=0$ for $m\neq0$.)
This is a manifestation of the ``equivalence theorem"
proven by Schwinger \cite{js} long ago. Though for $m_5=2m$
the value of $\th$ at $k=0$ is the same for pseudoscalar
and pseudovector couplings, we have seen that the pertinent
renormalization group trajectories are quite different in the two
cases.

\end{appendix}

\end{document}